\documentclass[twocolumn]{aastex62}
\usepackage{newtxtext,newtxmath}
\usepackage{ae,aecompl}
\usepackage{upgreek}
\usepackage{color}
\usepackage{url}
\usepackage{graphicx}
\usepackage{xspace}

\newcommand{\firstlong}{FIRST~J141918.9+394036\xspace}
\newcommand{\first}{FIRST~J1419+3940\xspace}

\graphicspath{{./}}

\received{XXX}
\revised{XXX}
\accepted{XXX}

\submitjournal{ApJL}

\shorttitle{Resolving a decades-long radio transient source}
\shortauthors{Marcote et al.}

\begin{document}

\title{Resolving the decades-long transient \firstlong: an orphan long gamma-ray burst or a young magnetar nebula?}

\correspondingauthor{B.~Marcote}
\email{marcote@jive.eu}

\author[0000-0001-9814-2354]{B.~Marcote}
\affiliation{Joint Institute for VLBI ERIC, Oude Hoogeveensedijk 4, 7991~PD Dwingeloo, The Netherlands}

\author{K.~Nimmo}
\affiliation{ASTRON, Netherlands Institute for Radio Astronomy, Oude Hoogeveensedijk 4, 7991~PD Dwingeloo, The Netherlands}
\affiliation{Anton Pannekoek Institute for Astronomy, University of Amsterdam, Science Park 904, 1098 XH Amsterdam, The Netherlands}

\author{O.~S.~Salafia}
\affiliation{Universit\`a degli Studi di Milano-Bicocca, Dip. di Fisica ``G. Occhialini'', Piazza della Scienza 3, 20126 Milano, Italy}
\affiliation{INAF - Osservatorio Astronomico di Brera, via E. Bianchi 46, 23807 Merate, Italy}
\affiliation{INFN - Sezione di Milano-Bicocca, Piazza della Scienza 3, 20126 Milano, Italy}

\author{Z.~Paragi}
\affiliation{Joint Institute for VLBI ERIC, Oude Hoogeveensedijk 4, 7991~PD Dwingeloo, The Netherlands}

\author{J.~W.~T.~Hessels}
\affiliation{ASTRON, Netherlands Institute for Radio Astronomy, Oude Hoogeveensedijk 4, 7991~PD Dwingeloo, The Netherlands}
\affiliation{Anton Pannekoek Institute for Astronomy, University of Amsterdam, Science Park 904, 1098 XH Amsterdam, The Netherlands}

\author{E.~Petroff}
\affiliation{Anton Pannekoek Institute for Astronomy, University of Amsterdam, Science Park 904, 1098 XH Amsterdam, The Netherlands}

\author{R.~Karuppusamy}
\affiliation{Max-Planck-Institut f\"ur Radioastronomie, Auf dem H\"ugel 69, D-53121 Bonn, Germany}


\begin{abstract}
    \citet{ofek2017} identified \firstlong (hereafter \first) as a radio source sharing similar properties and host galaxy type to the compact, persistent radio source associated with the first known repeating fast radio burst, FRB~121102. \citet{law2018} showed that \first is a transient source decaying in brightness over the last few decades. One possible interpretation is that \first is a nearby analogue to FRB~121102 and that the radio emission represents a young magnetar nebula (as several scenarios assume for FRB~121102).  Another interpretation is that \first is the afterglow of an `orphan' long gamma-ray burst (GRB). The environment is similar to where most such events are produced.  To distinguish between these hypotheses, we conducted radio observations using the European VLBI Network at 1.6\,GHz to spatially resolve the emission and to search for millisecond-duration radio bursts.  We detect \first as a compact radio source with a flux density of $620 \pm 20\,\mathrm{\upmu Jy}$ (on 2018 September 18) and a source size of $3.9 \pm 0.7\,\mathrm{mas}$ (i.e.\ $1.6 \pm 0.3\,\mathrm{pc}$ given the angular diameter distance of $83\,\mathrm{Mpc}$).  These results confirm that the radio emission is non-thermal and imply an average expansion velocity of $(0.10 \pm 0.02)c$. Contemporaneous high-time-resolution observations using the 100-m Effelsberg telescope detected no millisecond-duration bursts of astrophysical origin.  The source properties and lack of short-duration bursts are consistent with a GRB jet expansion, whereas they disfavor a magnetar birth nebula.
\end{abstract}



\section{Introduction} \label{sec:intro}

Very-long-baseline radio interferometric (VLBI) observations are a powerful way to study astrophysical transients because they provide milliarcsecond angular resolution imaging and astrometry.  Such transient events are produced by blast waves and slowly-evolving synchrotron afterglows, whose temporal evolution and interaction with the surrounding medium are well characterized by VLBI observations that can measure the projected size and proper motion of such emission.

For example, this technique was successfully used to spatially resolve the emission and measure the expansion speed of the afterglow associated with the long gamma-ray burst (GRB) 030329 \citep{pihlstrom2007}. VLBI observations have also been used to study the first detected binary neutron star merger, GRB~170817A \citep{abbott2017}. The obtained measurement of the proper motion and physical size constrained the nature of the source to be a relativistic jet \citep{mooley2018,ghirlanda2018}.
Furthermore, VLBI observations contributed to the first precise localization of a fast radio burst (FRB), the repeating source FRB~121102 \citep{spitler2014,spitler2016,scholz2016}. The burst source was associated with a compact ($< 0.7\,\mathrm{pc}$; \citealt{marcote2017}), persistent radio source with a luminosity of $\nu L_\nu \approx 3 \times 10^{38}\,\mathrm{erg\,s^{-1}}$ at 1.7~GHz \citep{chatterjee2017}, located inside a low-metallicity star-forming region in a dwarf galaxy at a redshift of $0.19273(8)$ \citep{tendulkar2017,bassa2017}.
The environment of FRB~121102 is remarkably similar to the ones where long GRBs (as well as superluminous supernovae) typically occur \citep{modjaz2008,metzger2017}, favoring several scenarios that consider repeating FRBs to be produced by newly-born magnetars created in such events \citep[see e.g.][]{margalit2018,piro2018}. FRBs could thus be detectable at the sites of long GRBs, and the persistent source associated with FRB~121102 could be the longer-lived nebula following the afterglow of one of these events. In any case, FRBs are expected to be produced in relatively young objects ($\sim 10$--$100\,\mathrm{yr}$) with possibly associated radio nebulae \citep{murase2016,kashiyama2017,omand2018}.

Based on the properties of FRB~121102's persistent radio source and host galaxy, \citet{ofek2017} identified a number of similar sources in the Very Large Array (VLA) FIRST catalogue.
\citet{law2018} showed that one of these sources, \firstlong (hereafter \first) is a slowly declining transient. Using archival observations, they showed that the source declined from $\sim 26\,\mathrm{mJy}$ (at 1.4~GHz) in 1993 to $\lesssim 0.4\,\mathrm{mJy}$ (at 3~GHz) in 2017.
Both the light-curve and the inferred luminosities of $\nu L_\nu \gtrsim 3 \times 10^{38}\,\mathrm{erg\,s^{-1}}$ are consistent with the afterglow of a long GRB, requiring a released kinetic energy of $\sim 10^{51}\,\mathrm{erg}$ at the time of the explosion (estimated to be $\sim 25$--$30\,\mathrm{yr}$ ago).  No convincing association with a previously detected GRB could be made, however \citep{law2018}.

\first is associated with a small star-forming galaxy at redshift of $z \approx 0.01957$. Both sources, \first and FRB~121102, show similar environments: both show compact and persistent radio emission with luminosities of $\sim 10^{38}\,\mathrm{erg\ s^{-1}}$ located inside star-forming regions with equivalent star formation rates in similar sized dwarf galaxies. Their physical nature could thus also be similar, and \first might be associated with a source capable of producing FRBs.

Here we present European VLBI Network (EVN) radio observations of \first that provide the first constraints on the source compactness, coupled with simultaneous searches for millisecond-duration bursts. We present the observations and data reduction in Section~\ref{sec:observations}. We describe the results in Section~\ref{sec:results}, and their implications for the nature of \first in Section~\ref{sec:interpretation}. Finally, we present our conclusions in Section~\ref{sec:conclusions}.

\section{Observations and data reduction} \label{sec:observations}

We observed \first on 2018 September 18 between 12:00 and 19:00~UTC at 18~cm (1.6~GHz) with the EVN, involving a total of 12 stations: Jodrell Bank Mark2, Westerbork single-dish, Effelsberg, Medicina, Onsala 25-m, Tianma, Toru\'{n}, Hartebeesthoek, Sardinia, and three stations from e-MERLIN (Cambridge, Defford, and Knockin). The data were recorded with a total bandwidth of 128~MHz, and correlated in real time (e-EVN operational mode) at JIVE (The Netherlands) using the SFXC software correlator \citep{keimpema2015}. The data were divided into eight subbands of 64 channels each, with full circular polarization products, and 1-s time averaging.
We also buffered the baseband EVN data in parallel so that high-time-resolution correlations could be produced afterwards, if a millisecond-duration radio burst was detected.

Furthermore, we simultaneously observed \first in the frequency range $1580$--$1736\,\mathrm{MHz}$ using the $100$-m Effelsberg telescope and the PSRIX pulsar data recorder \citep{lazarus2016}. We recorded with two summed linear polarizations, achieving a gain of $1.5$~K~Jy$^{-1}$ and a receiver temperature of $25$~K. The total bandwidth of $156\,\mathrm{MHz}$ was divided into $10$ subbands --- each one further divided into $64$ channels and recorded with $32$-bit time samples.  The ultimate time and frequency resolution of the data were $40.96\,\mathrm{\upmu s}$ and $0.2438\,\mathrm{MHz}$, respectively. Before processing, the subbands were combined into a single band and the data were converted to $4$-bit samples to ensure compatibility with the {\tt PRESTO} pulsar analysis software suite \citep{ransom2001thesis}.

\subsection{Interferometric data}

We observed J1642+3948 as fringe finder and J1419+3821 (located at only $1.3\degr$ from \first) as phase calibrator. We scheduled a phase-referencing cycle of 4.5~min on the target and 1.5~min on the phase calibrator, achieving a total time of $\sim 4.5\,\mathrm{h}$ on \first.

The interferometric data were reduced using {\tt AIPS}\footnote{The Astronomical Image Processing System ({\tt AIPS}) is a software package produced and maintained by the National Radio Astronomy Observatory (NRAO).} \citep{greisen2003} and {\tt Difmap} \citep{shepherd1994} following standard procedures. A-priori amplitude calibration was performed using the known gain curves and system temperature measurements recorded individually on each station during the observation.
We used nominal system equivalent flux density (SEFD) values for the following stations: Jodrell Bank Mark2, Tianma, and the e-MERLIN stations.
We manually flagged data affected by radio frequency interference (RFI) and then we fringe-fitted and bandpass-calibrated the data using the fringe finder and the phase calibrator. We imaged and self-calibrated the phase calibrator in {\tt Difmap} to improve the final calibration of the data. The obtained solutions were then transferred to the target, which was subsequently imaged.

\begin{figure}
    \begin{center}
        \includegraphics[width=\columnwidth]{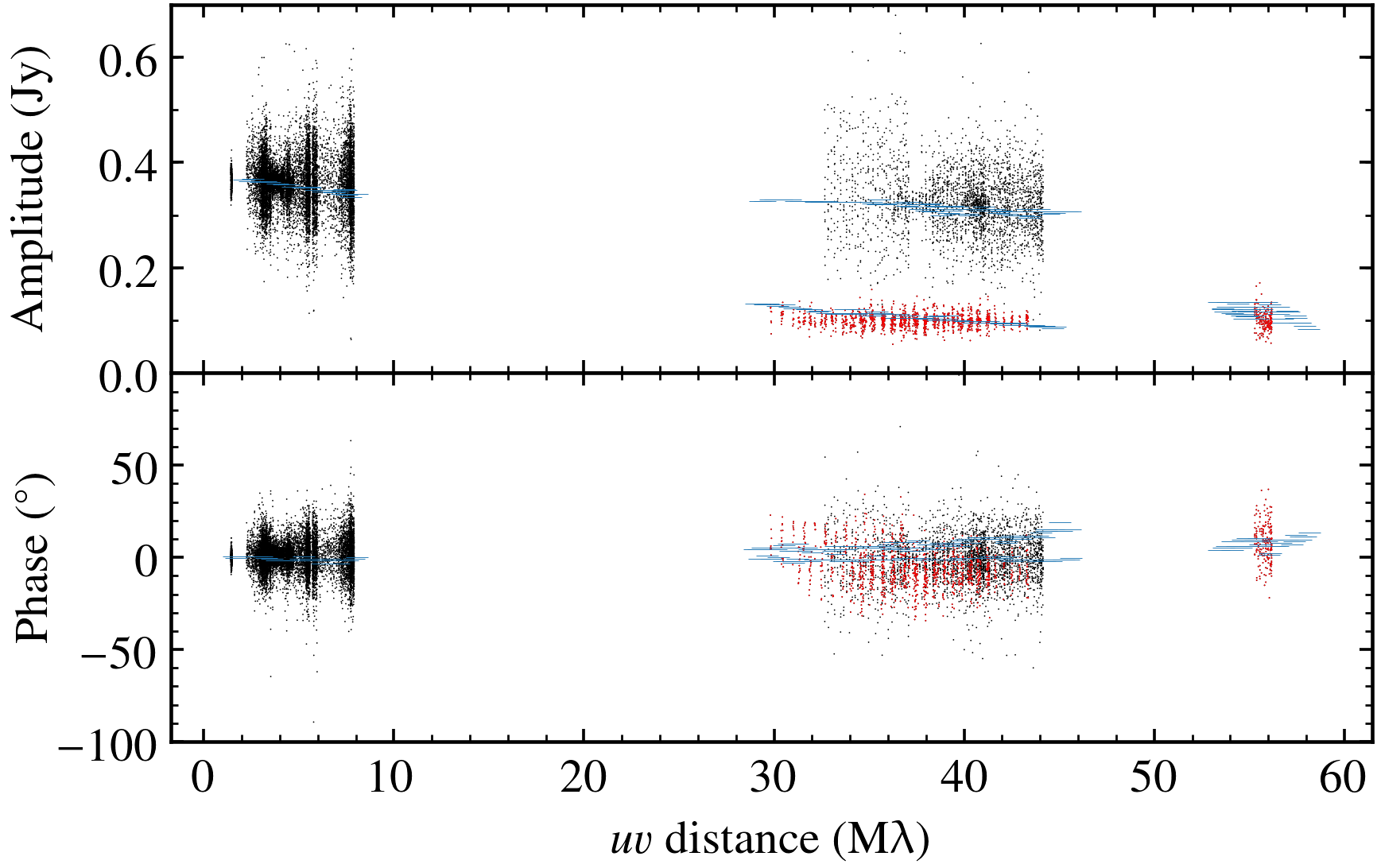}\\[+10pt]
        \includegraphics[width=\columnwidth]{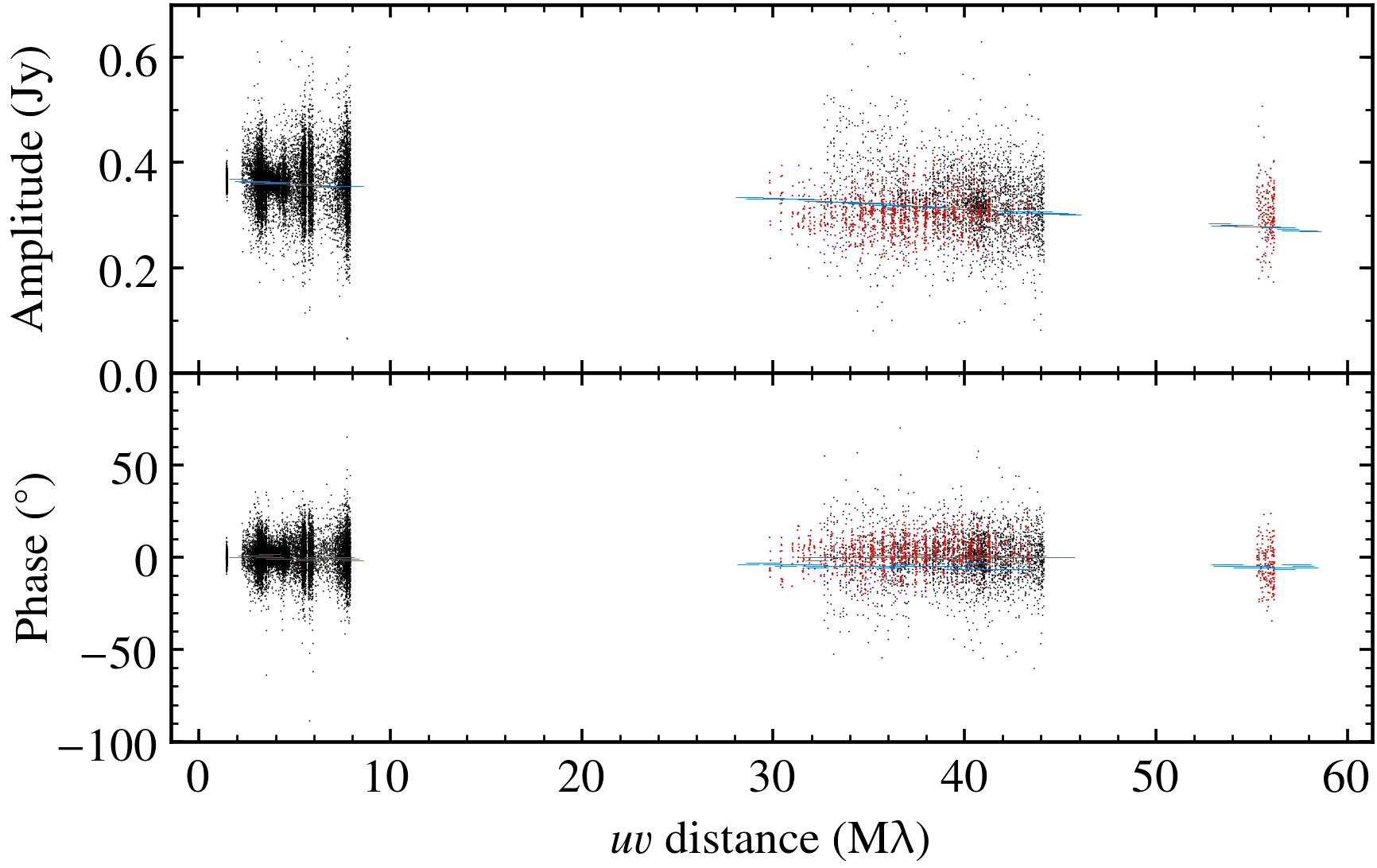}
        \caption{The obtained visibility data (amplitudes and phases) for the phase calibrator source, J1419+3821, after the original calibration (top) and in the alternate calibration where the gain calibration of the Tianma station was calibrated using a source model from only the stations with a robust amplitude calibration (bottom). See Sect.~\ref{sec:observations} for details. Red dots represent data from the baselines including the Tianma station. Blue lines represent the source model in each case.}
        \label{fig:radplot}
    \end{center}
\end{figure}

%

We note that Tianma did not produce reliable system temperature values during the experiment. Most of these measurements failed and the existing ones exhibited a much larger scatter than usual. Therefore we used the nominal SEFD for amplitude calibration. This typically produces a satisfactory a-priori calibration that can be further improved during imaging and self-calibration. Tianma, however, provides the longest East-West baselines in our array with no equivalent baselines to compare with, and it also does not have short spacings to establish a reliable station calibration. In this case imaging and parametrization of source properties by model-fitting is complicated due to the fact that some source parameters may correlate with the Tianma station gain \citep{natarajan2017}.

Figure~\ref{fig:radplot} displays the visibility amplitudes and phases as a function of projected baseline length in units of observing wavelength. The top panel shows the initial calibration, with the Tianma data highlighted in red. The low amplitudes may be consistent with a source that is very compact in general, but well resolved in the East-West direction. The bottom panel shows the data after we apply an amplitude correction factor for Tianma, based on a source model obtained by only using the stations with robust calibration. The required scaling factor was about three, implying that the station could have been much less sensitive than expected.

Due to the uncertainty with the Tianma calibration, we decided to take the following procedure to analyze the data. Instead of fitting an elliptical-Gaussian model brightness distribution to the $uv$-data in model fitting, we assumed a circular-Gaussian brightness distribution. This is expected to be less sensitive to uncertainties in station gain calibration \citep{natarajan2017}. In addition, we looked at the results derived from the following cases: Tianma removed from the data set, Tianma present with nominal gain calibration, and Tianma present but with its gain scaled to be in agreement with the most compact possible solution (as explained above). As we will see in Sect.~\ref{sec:results}, the fitted source sizes differ somewhat, but in all cases they support the same main conclusion: that our target is resolved on milliarcsecond scales.

\subsection{High-time resolution data}

The high-time-resolution Effelsberg data were analyzed to search for individual millisecond bursts or a periodic signal. First, using {\tt PRESTO}'s {\tt rfifind}, we identified specific time samples and frequency channels contaminated by RFI. The regions highlighted by {\tt rfifind} and the frequency range $1610$--$1631\,\mathrm{MHz}$, associated with RFI from the Iridium satellites, were masked prior to conducting the analysis. We then dedispersed the $4$-bit data using the {\tt PRESTO} tool {\tt prepsubband} for $2500$ trial dispersion measures (DMs) in the range $0$--$1210.8\,\mathrm{pc\,cm^{-3}}$. The resulting dedispersed time series were then searched for single pulses above a $6$-$\sigma$ threshold using {\tt PRESTO}'s {\tt single\_pulse\_search.py}, which applies a matched-filter technique using boxcar functions of various widths, and in our search was sensitive to burst durations in the range $40.96\, \mathrm{\upmu s}$ and $0.02\,\mathrm{s}$. Dynamic spectra of the identified single-pulse candidates were generated and inspected by eye to distinguish between astrophysical signals and RFI.

In addition, a Fourier-domain search was performed on each individual dedispersed time series using {\tt PRESTO}'s {\tt accelsearch}, in order to search for periodic signals. Potential periodic signals were sifted using {\tt ACCEL\_sift.py}, and the remaining candidates were inspected by eye after folding using {\tt prepfold}.

The RFI mitigation process was unable to remove all instances of RFI in the data. We calculate that of the $\sim4.3$-$\mathrm{h}$ Effelsberg on-source time, approximately $92.4\%$ was examined for bursts and periodic signals. Note the discrepancy between the Effelsberg on-source time ($\sim 4.3\,\mathrm{h}$) and the EVN on-source time ($\sim 4.5\,\mathrm{h}$) which is due to Effelsberg's longer slew time compared with other antennas.

The aforementioned analysis strategy was verified using similar data targeting pulsar PSR~B2020+28. We performed a blind search and detected both individual pulses and the known periodicity of this pulsar.

\section{Results and discussion} \label{sec:results}

\subsection{On the persistent emission}

\begin{figure}
    \begin{center}
        \includegraphics[width=\columnwidth,trim={0 0 1.0cm 0}]{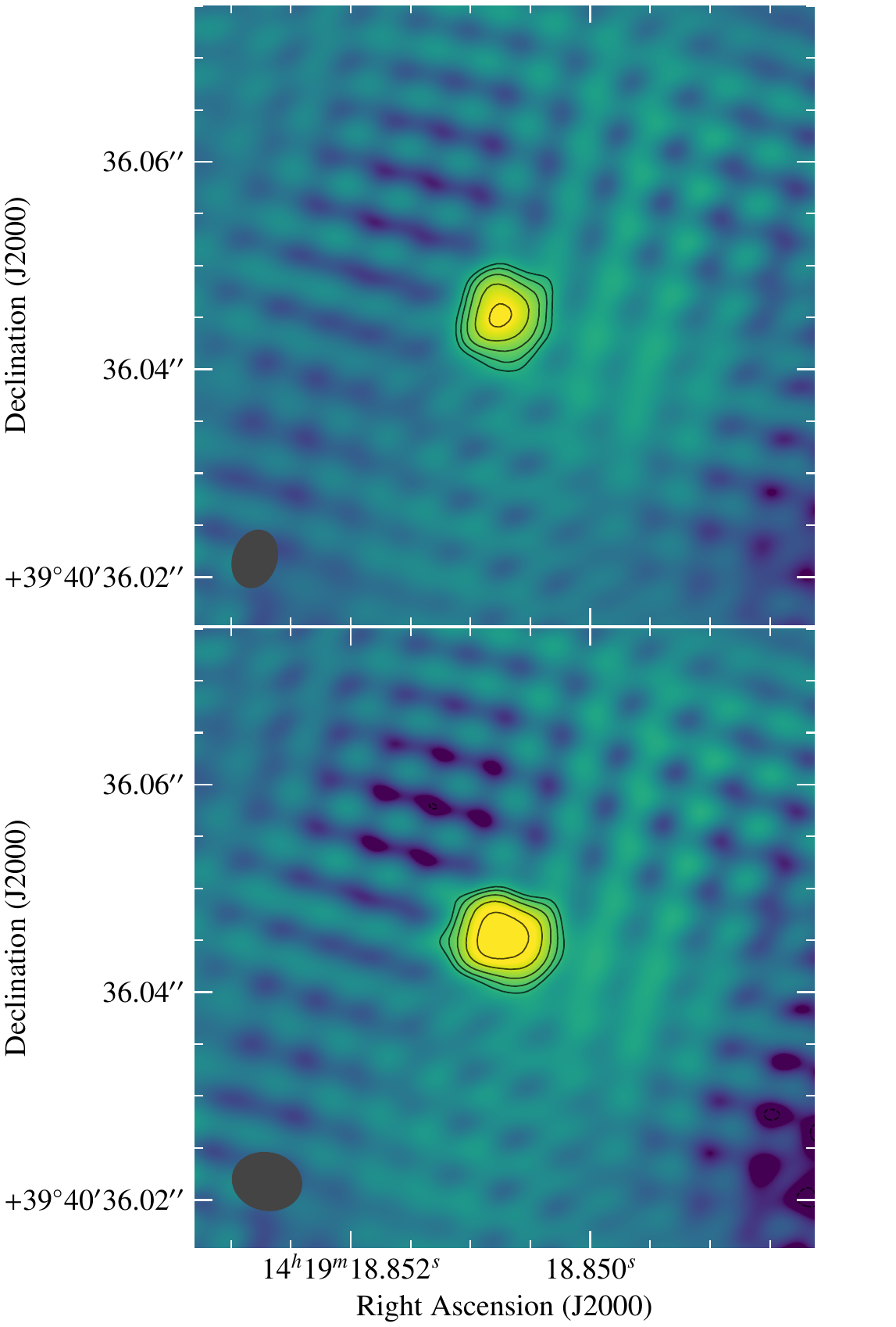}
        \caption{Images of \first at 1.6~GHz with the EVN on 2018 September 18 derived from the two gain calibrations performed on Tianma data (original, top, and scaled, bottom). Contours start at a 3-$\sigma$ rms noise level of $25$ and $29\,\mathrm{\upmu Jy\,beam^{-1}}$, respectively, and increase by factors of $\sqrt{2}$. The synthesized beams are represented by the dark gray ellipses at the bottom left corner of each image.}
        \label{fig:image}
    \end{center}
\end{figure}

\begin{figure}
    \begin{center}
        \includegraphics[width=\columnwidth]{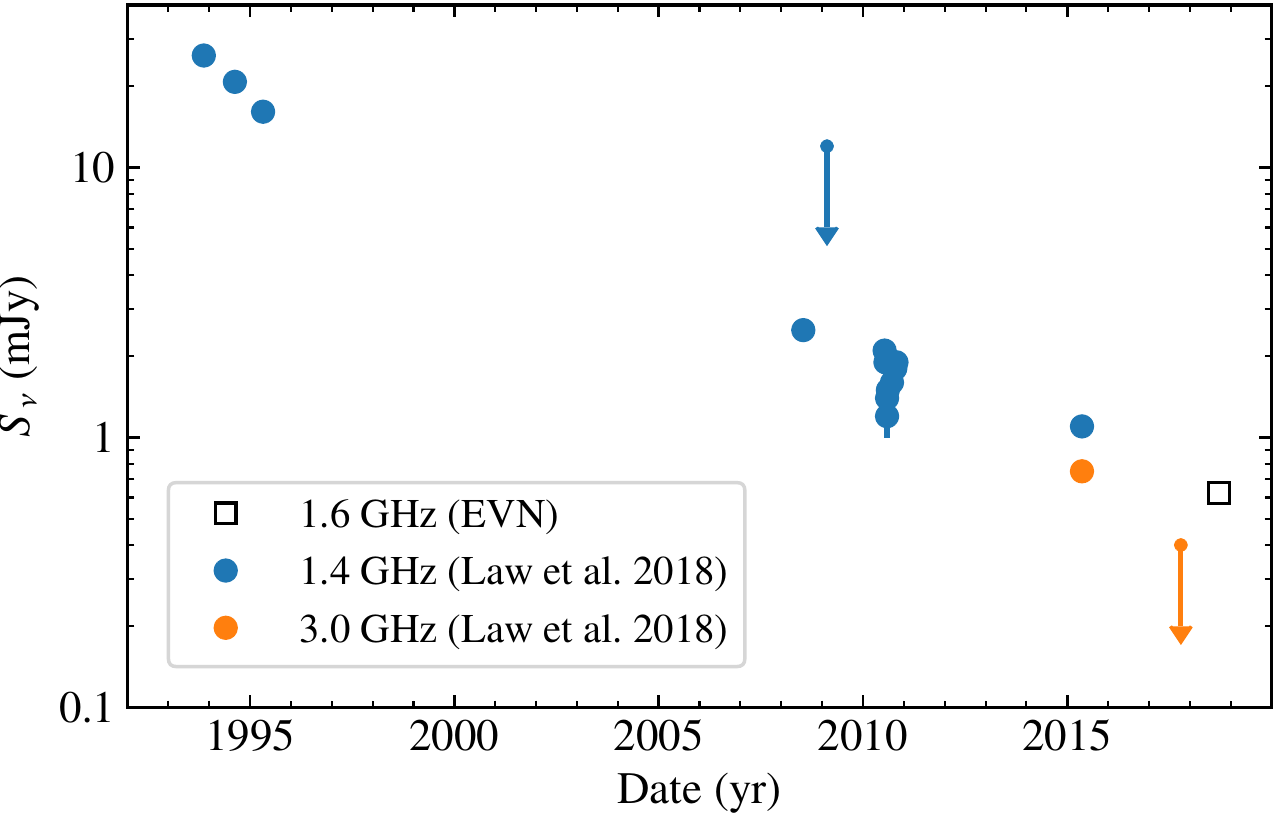}
        \caption{Light-curve of \first during the last 25~yr at 1.4--1.6~GHz (blue circles and open square) and 3.0~GHz (orange circles). Errors bars represent 1-$\sigma$ uncertainties (hidden by the size of the markers in most cases). Arrows represent 3-$\sigma$ upper-limits.}
        \label{fig:lightcurve}
    \end{center}
\end{figure}

\first is detected on 2018 September 18 as a radio source that is compact on milliarcsecond scales (see Fig.~\ref{fig:image}), with a flux density of $620 \pm 20\,\mathrm{\upmu Jy}$ at a position of:
\begin{align*}
    \alpha (\text{J2000}) &= 14^{\rm h}19^{\rm m}18.850722^{\rm s}\quad \pm 0.23\,\text{mas}\\
    \delta (\text{J2000}) &= 39\degr 40^{\prime} 36.04520^{\prime\prime} \phantom{\;\;}\quad \pm 0.23\,\text{mas},
\end{align*}
where the quoted uncertainties represent the 1-$\sigma$ confidence interval and take into account the statistical uncertainties in the image (0.2~mas in both $\alpha$ and $\delta$), the uncertainty in the phase calibrator position \citep[0.1~mas;][]{beasley2002,gordon2016}, and the estimated uncertainties associated with the phase referencing technique \citep[0.06 and 0.04~mas for $\alpha$ and $\delta$, respectively;][]{pradel2006}.
The obtained position is consistent with the one reported from the FIRST survey \citep{law2018}, as well as the preliminary results published in \citet{marcote2018atel}. The measured flux density on 2018 September 18 follows the declining trend of the light-curve reported from observations with the Karl G.\ Jansky Very Large Array, VLA (see Fig.~\ref{fig:lightcurve}). 
Given the luminosity distance of 87~Mpc, the obtained flux density corresponds to an isotropic luminosity $\nu L_\nu = (9.4 \pm 0.3) \times 10^{36}\,\mathrm{erg\,s^{-1}}$. Together with the last published VLA observation at 3.0~GHz, and considering the same value for our epoch (i.e.\ no declining trend is assumed), we can place a conservative 3-$\sigma$ upper limit on the spectral index between 1.6 and 3.0~GHz of $\alpha \lesssim -0.65$ (where $S_\nu \propto \nu^\alpha$).

\begin{deluxetable*}{lccccc}
    \tablecaption{Properties of \first measured following different imaging approaches.
    \label{tab:data}}
    \tablehead{
        & \colhead{rms} & \colhead{Peak brightness} & \colhead{Flux Density} & \colhead{size} & \colhead{synthesized beam}\\[-4pt]
        & \colhead{(${\rm\upmu Jy\, beam^{-1}}$)} & \colhead{(${\rm\upmu Jy\, beam^{-1}}$)} & \colhead{(${\rm\upmu Jy}$)} & \colhead{(mas)} & \colhead{(mas $\times$ mas,\ $\degr$)}
    }
    \startdata
        Default calibration & $19$ & $300$ & $620 \pm 20$ & $4.3 \pm 0.8$ & $5.5 \times 4.1,\ -18\degr$ \\
        Without Tianma & $30$ & $510$ & $630 \pm 30$ & $3.9 \pm 0.9$ & $24 \times 5.3,\ 75\degr$ \\
        Corrected Tianma & $25$ & $459$ & $620 \pm 30$ & $3.4 \pm 0.7$ & $6.6 \times 5.4,\ 78\degr$\\
    \enddata
\end{deluxetable*}

\first is significantly resolved in the obtained images given the size of the synthesized beam ($5.5 \times 4.1\,\mathrm{mas^2}$), as the measured size is larger than the minimum resolvable size by the array \citep[see][for a detailed explanation]{martividal2012, natarajan2017}. By fitting a circular Gaussian to the $uv$ data we measure a source size of $4.3 \pm 0.8\,\mathrm{mas}$, where the uncertainty has been estimated through a $\chi^2$ test. However, we note that the gain calibration of Tianma constitutes a potential source of systematic errors in the size measurement. To provide a more reliable measurement, we produced images without this station. Despite having poorer resolution (synthesized beam of $24 \times 5.3\,\mathrm{mas^2}$), we obtained a size that is significant and consistent within uncertainties with the value quoted above ($3.9 \pm 0.9\,\mathrm{mas}$). Finally, we imaged the source with the gain correction applied to Tianma (as mentioned in the previous section), which provides the most stringent lower limit on the source size (i.e.\ assuming a point-like source during calibration).
In this case we measured a source size of $3.4 \pm 0.7\,\mathrm{mas}$. The contribution of the longest baselines is therefore not critical as we obtain consistent results from all cases. We summarize the results of these different analyses in Table~\ref{tab:data}.

For comparison, the phase calibrator, J1419+3821, exhibits a main compact component with a measured size of $1.1$--$2.9\,\mathrm{mas}$ in all cases. The fact that the measured sizes are siginificantly different -- while they are seen along almost the same Galactic line of sight -- means that they are most likely intrinsic sizes, rather than due to scatter broadening. At the high Galactic latitudes of $\sim 67^{\circ}$ scatter broadening at GHz frequencies is almost negligible, and one would not expect strong variations of scattering size on small angular scales either \citep[see e.g.][]{pushkarev2015}.

We thus conclude that \first is significantly resolved, with an angular size of $3.9_{-0.5}^{+0.4}\phantom{}_{-0.2}^{+0.3}\,\mathrm{mas}$, where the first uncertainties take into account the dispersion of the values from the different analyses, and the second ones consider the estimated statistical uncertainties on the value. Given that the angular diameter distance to the source is 83~Mpc \citep{law2018}, we derive a projected physical size of $1.6 \pm 0.3\,\mathrm{pc}$. This size also implies a brightness temperature of $T_b \sim 1.1 \times 10^7\,\mathrm{K}$, which clearly points to a non-thermal origin for the emission.

\citet{law2018} estimated that the putative GRB producing the observed afterglow likely took place around $\sim 25$--$30\,\mathrm{yr}$ ago. Considering an estimated central date for the explosion of $\sim 1993$ and taking into account the given uncertainties, the afterglow must have a mean expansion velocity of $v = (3.0 \pm 0.6) \times 10^4\,\mathrm{km\,s^{-1}}$, or $(0.10 \pm 0.02) c$, consistent with a mildly relativistic expansion. We note that the calculated expansion velocity is an average over the whole lifetime, during which a significant deceleration has likely occurred.

\subsection{On the single burst searches}\label{sec:singleburstsearch}

We detected no astrophysical single pulses or periodic signals in the high-time-resolution Effelsberg data. We can estimate the expected dispersion measure (DM) towards \first using Galactic electron density models (NE2001; \citealt{cordes2001}, YMW16; \citealt{yao2017}). For an extragalactic source, the observed DM can be divided into four components along the line of sight:
\begin{equation}
    \label{eq:dm}
    \mathrm{DM}_{\rm obs} = \mathrm{DM}_{\rm MW} + \mathrm{DM}_{\rm MW_{\rm halo}} + \mathrm{DM}_{\rm IGM} + \mathrm{DM}_{\rm host}.
\end{equation}
The Milky Way contribution to the DM along the line of sight is divided into the disk and spiral arm component, ${\rm DM}_{\rm MW}$, and the Galactic halo component, ${\rm DM}_{\rm MW_{\rm halo}}$. The former, ${\rm DM}_{\rm MW}$, is $44$ and $39\,\mathrm{pc\,cm^{-3}}$ calculated using the NE2001 and YMW16 models, respectively. The uncertainties in these contributions are not well quantified, but are likely on the order of $20\%$. Using this, we can derive an approximate range of: $30 \lesssim {\rm DM}_{\rm MW} \lesssim 50\,\mathrm{pc\,cm^{-3}}$. We apply a Galactic halo contribution of $\sim60$--$ 100\,\mathrm{pc\,cm^{-3}}$ to the DM \citep{prochaska2019}. Given that the redshift of \first is $0.01957$ \citep{law2018}, the mean intergalactic medium (IGM) contribution to the DM is ${\rm DM}_{\rm IGM}\simeq 20\,\mathrm{pc\,cm^{-3}}$ \citep{ioka2003, inoue2004}. We assume that the DM contribution of the host galaxy of \first, ${\rm DM}_{\rm host}$, is comparable to that of the host galaxy of FRB~121102: $55 \lesssim {\rm DM}_{\rm host} \lesssim 225\,\mathrm{pc\,cm^{-3}}$ \citep{tendulkar2017}.
Combining all individual components using equation \eqref{eq:dm} results in the approximate range $160 \lesssim {\rm DM}_{\rm obs} \lesssim 400\,\mathrm{pc\,cm^{-3}}$.

From the single pulse candidates reported using {\tt single\_pulse\_search.py}, an astrophysical burst would be identifiable provided the signal-to-noise ratio exceeds $\sim 10$. We can estimate the fluence limit of our search using
\begin{equation}
    F = (S/N)_{\rm min}\frac{T_{\mathrm{sys}}}{G}\sqrt{\frac{W_{b}}{n_{\mathrm{pol}}\ \Delta \nu}}
\end{equation}
\citep[following][]{Cordes2003}, where $(S/N)_{\rm min}$ is our detection threshold of $10$, $T_{\mathrm{sys}}$ is the system temperature, $G$ is the telescope gain, $n_{\rm pol}$ is the number of recorded polarizations, $\Delta \nu$ is the total bandwidth, and $W_{b}$ is the observed width of the burst. The observed width, $W_{b}$, accounts for broadening of the intrinsic width due to the finite time sampling of the data, intra-channel smearing, smearing due to DM-trial spacing, and scatter broadening. FRB~121102 has been shown to exhibit individual bursts with widths $\lesssim 30\,\mathrm{\upmu s}$ \citep{michilli2018} and there have been observations of FRBs with widths as large as $\sim 30\,\mathrm{ms}$ \citep{petroff2016}\footnote{All published FRBs and their properties can be found in the FRB Catalogue: {\tt http://www.frbcat.org}.}. Taking a DM of $300\,\mathrm{pc\,cm^{-3}}$ and intrinsic widths $30\,\mathrm{\upmu s}$--$30\,\mathrm{ms}$, we find our fluence limit ranges from $0.1\,\mathrm{Jy\,ms}$ to $8\,\mathrm{Jy\,ms}$.
\begin{figure}
\centering
\includegraphics[width=\columnwidth]{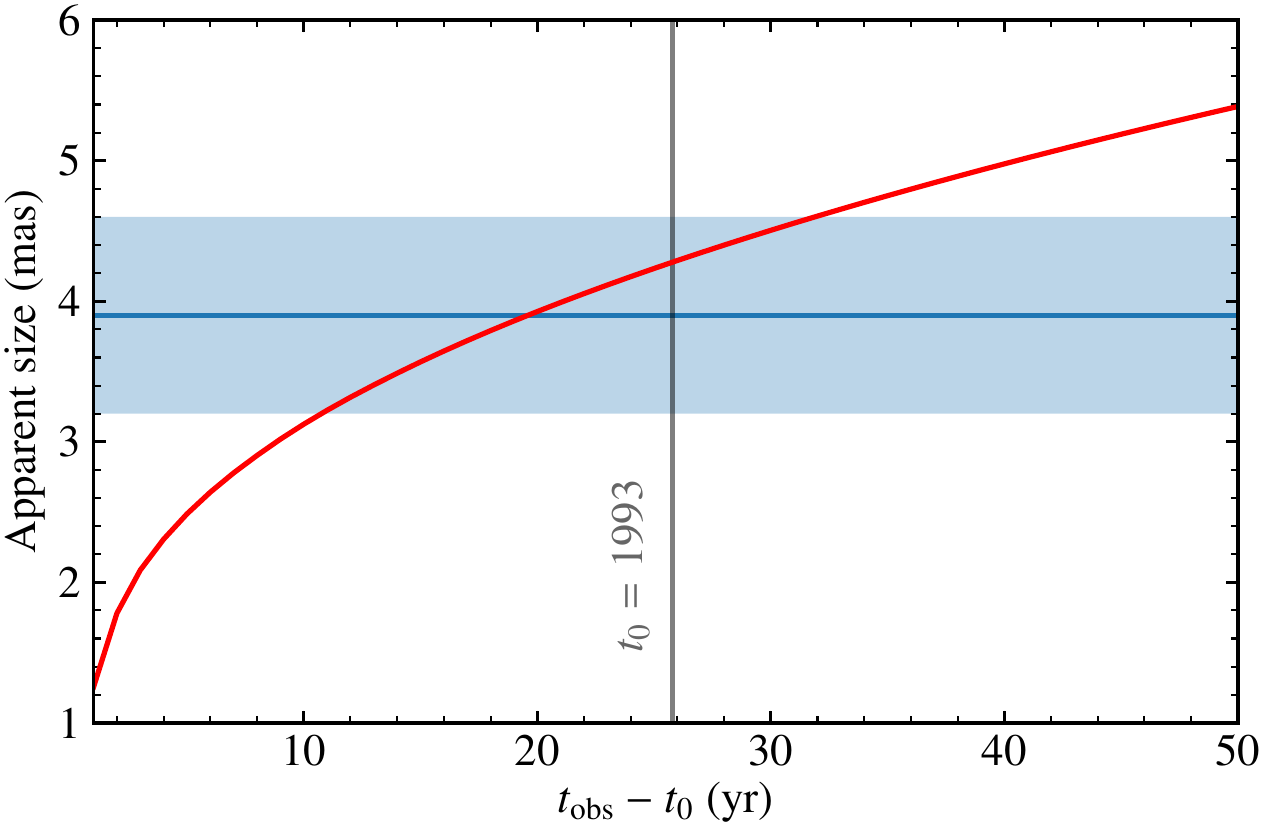}
\caption{Apparent source size evolution. The red line shows the predicted apparent size evolution for a jet with parameters as those proposed by \citet{law2018}. The blue line and the lighter blue band show our measured apparent size and its 1-$\sigma$ uncertainty of $\theta_\mathrm{s}= 3.9 \pm 0.7\,\mathrm{mas}$. The grey vertical line marks the source age at the time of our observation, assuming that it originally exploded in 1993.}
\label{fig:size_predictions}
\end{figure}

\section{Interpretation} \label{sec:interpretation}

\subsection{Measured source size}

In order to compare our measurements with the scenarios proposed by \citet{law2018}, we need to compute the expected apparent size of the source. At the time of our observation, $t_\mathrm{obs}\sim 30\,\mathrm{yr}\sim 10^4\,\mathrm{d}$ after the initial explosion, the external shock produced by the GRB jet upon deceleration into the interstellar medium (ISM) is expected to be non-relativistic, and to have become essentially spherical. Detailed, long term numerical relativistic hydrodynamics simulations \citep{Zhang2009} indeed show that the blast wave is well-described by the spherical, non-relativistic Sedov-Von Neumann-Taylor solution after a time $5 t_\mathrm{NR}\approx 2\times 10^3 E_\mathrm{iso,53}^{1/3}\,n_\mathrm{1}^{-1/3}\,\mathrm{d}$, where $t_\mathrm{NR}$ is the light-crossing time of the Sedov length associated to the jet.  The initial relativistic expansion phase, however, can still have effects on the relation between observed time and projected size. For that reason, we  compute the jet deceleration dynamics and spreading employing the ``trumpet'' model from \citet{Granot2012}, which has been shown to be in good quantitative agreement with results from numerical relativistic hydrodynamics simulations. The observed size is estimated as the maximum projected size of the equal-arrival-time surface, relativistic beaming of radiation being negligible in our late-time observations. Using the same jet parameters as \citet{law2018}, namely an isotropic equivalent energy $E_\mathrm{iso} = 2\times 10^{53}\,\mathrm{erg}$, an ISM number density $n=10\,\mathrm{cm}^{-3}$ and a viewing angle $\theta_\mathrm{v}=0.6\,\mathrm{rad}$, and further assuming a jet half-opening angle $\theta_\mathrm{j}=0.1\,\mathrm{rad}$ (which implies a total jet energy $E_\mathrm{jet}\sim 10^{51}\,\mathrm{erg}$), we obtain the size evolution shown by the red solid line in Fig.~\ref{fig:size_predictions}, which is fully compatible with the measured one, assuming that the GRB took place in $\sim 1993$. This disfavors the alternative scenario of a magnetar birth nebula, which would exhibit a significantly smaller size ($\lesssim 0.1\,\mathrm{pc}$), due to the much lower expansion velocity \citep{murase2016}.

\subsection{Flux density}

While the measured size agrees well with the GRB scenario proposed by \citet{law2018}, our  measured flux density $S_\mathrm{1.6\,GHz}= 620 \pm 20\,\mathrm{\upmu Jy} $ is low when compared to the extrapolation of their model. More precisely, adopting the same assumptions as \citet{law2018}, namely quasi-isotropic, adiabatic expansion, Deep Newtonian regime and an electron power law index $p = 2.2$, the flux density should follow $S_\nu \propto \nu^{-0.6}t^{-0.96}$. Using the latest VLA detection as reference, which yielded a flux density of $1.1\pm0.1\,\mathrm{mJy}$ at $1.52\,\mathrm{GHz}$ on 2015 May 11, and assuming the GRB to have happened in 1993, we should have measured $S_\mathrm{1.6\,GHz} \approx 930\pm 85\,\mathrm{\upmu Jy}$ at the time of our observation, which is $\sim 3.5\sigma$ (summing the uncertainties in quadrature) above our measured flux density. As noted by \citet{law2018}, the latest VLASS non-detection $S_\nu < 400\,\mathrm{\upmu Jy}$ at $3\,\mathrm{GHz}$ on 2017 October 11 already pointed to a faster decline after 2015. Several physical processes could lead to a steepening in the decay of the lightcurve, e.g.:

\begin{itemize}
    \item The conditions in the shocked fluid could be changing as a consequence of the transition to the non-relativistic, Deep Newtonian phase, e.g.~the fraction $\epsilon_\mathrm{e}$ of shock energy given to electrons could decrease, or the electron momentum distribution power law index $p$ could decrease from $p = 2.2$ towards $\sim 2$ \citep{Sironi2013}. Both these effects would result in a steepening of the flux decay;
    \item Contrary to what is stated by \citet{law2018}, the steepening could also be due to the shock crossing a dip in the ISM density. According to the argument by \citet{law2018}, based on \citet{Nakar2007} and \cite{Mimica2011}, an ISM density drop would result only in a smooth, slow change in the lightcurve. This is essentially a consequence of the assumption that the shock is relativistic ($\Gamma \gg 1$), in which case the angular time scale $R/2\Gamma^2 c$ would be of the same order as the observer time $t_\mathrm{obs}$, and therefore any change in the shock conditions would be smeared out over that time scale. In our case, conversely, the shock expansion speed is non-relativistic, and the angular time scale is $\sim R/2c \ll t_\mathrm{obs}$ (using our size measurement, we have $R/c\sim 2.5\,\mathrm{yr}$, which is significantly smaller than the explosion age $t_\mathrm{obs}\gtrsim 25\,\mathrm{yr}$), so that a drop in the ISM density at a radius slightly smaller than the observed size $\sim 1.6\,\mathrm{pc}$ may justify the flux deficit. Such a drop could mark the outer radius of the star-forming region where the GRB exploded. Let us caution, though, that this requires some fine-tuning. In order for the shock to entirely cross the outer edge of the star-forming region in a short enough time, the latter should be approximately spherical and nearly concentric to the shock. By using the same shock dynamics model as in the previous section, we estimate the current shock expansion velocity to be $v_\mathrm{s}\sim 0.03c \approx 9\,000\,\mathrm{km\,s^{-1}}$. The shock thus travelled a distance $\Delta R\sim 0.03\,\mathrm{pc}$ between the latest VLA detection at $1.5\,\mathrm{GHz}$ and our observation. The centers of the shock wave and the star-forming region, assumed spherical, should therefore be located less than $\Delta R$ away from each other. Since the flux deficit we observe amounts to a factor $\sim 2/3$ reduction with respect to the expected value, we can partially relax these requirements, allowing the outer edge of the star-forming region to have some structure -- e.g.~bumps, filaments, a non-spherical shape -- as long as $\sim 1/3$ (in terms of solid angle) of the shock wave still experiences a sharp density drop in the required time. This leads us to conclude that, while the arguments against an ambient medium density drop proposed by \cite{law2018} do not hold in this case, this kind of explanation for the flux variation, while not impossible, remains rather unlikely.
\end{itemize}

Finally, we note that the flux deficit cannot be understood as due to scintillation-induced fluctuations, as the apparent size of the source is too large \citep{Goodman1997}.

\subsection{Comparison with FRB~121102} \label{sec:frb121102host}

The association of FRB~121102 with a persistent radio counterpart \citep{chatterjee2017,marcote2017} led to the discovery of \first, the characteristics of which match those of the persistent source coincident with FRB~121102 \citep{ofek2017,law2018}: i.e., a compact radio source with a similar luminosity and co-located with a star-forming region of a dwarf galaxy. It is, therefore, natural to compare \first with the radio counterpart of FRB~121102.

The declining light curve of \first contrasts with the persistent emission from FRB~121102 \citep[Plavin et al.\ in prep]{chatterjee2017}. Another discrepancy that arises is that the obtained source size of \first is significantly larger than the one associated with FRB~121102 ($< 0.7\,\mathrm{pc}$; \citealt{marcote2017}), implying a much higher expansion velocity. These differences can, naturally, be explained by a younger age of \first ($\sim 30\,\mathrm{yr}$) when compared with FRB~121102 ($\sim 100\,\mathrm{yr}$; \citealt{metzger2017,piro2018}). We note that although we conclude our observations are consistent with GRB jet expansion, we do not rule out the presence of a nebula driven by a highly magnetized neutron star contributing to a fraction of the radio emission observed. The presence of such nebula could cause the light curve to plateau at late-times.

It has been hypothesized that the birth of a millisecond magnetar can connect FRB~121102 with long GRBs or superluminous supernovae \citep[SLSNe;][]{metzger2017}. If we assume that millisecond magnetars can produce FRBs similar to what is observed in FRB~121102, and that a magnetized neutron star resides within the radio source \first, we might expect FRBs from this source. The comparable ages of \first and FRB~121102 leads to our assumption that the compact object residing within \first is emitting bursts with a comparable energy distribution and duty cycle to FRB~121102.

Bursts from the repeating FRB~121102 have been observed with fluences of $\sim 0.02\,\mathrm{Jy\,ms}$ \citep{gajjar2018} to $\gtrsim 7\,\mathrm{Jy\,ms}$ \citep{marcote2017} and widths ranging from $\lesssim 30\,\mathrm{\upmu s}$ \citep{michilli2018} to $\sim 8.7\,\mathrm{ms}$ \citep{spitler2016}. Taking this range of fluence values at the luminosity distance of FRB~121102 ($972\,\mathrm{Mpc}$) and scaling to the luminosity distance of \first ($87\,\mathrm{Mpc}$), gives an estimated fluence range of $2.5$--$870\,\mathrm{Jy\,ms}$. For bursts with widths exceeding $\sim 9\,\mathrm{ms}$, the fluence limit of our search increases beyond $2.5\,\mathrm{Jy\,ms}$ (see Section \ref{sec:singleburstsearch}). Under the assumption that \first is producing bursts with widths comparable to that of FRB~121102 and with an alignment (with respect to the observer) consistent with FRB~121102, single bursts from this source would be identifiable in the data.


The lack of short-duration bursts in our observation could imply that the source is in a quiescent state, similar to the behaviour observed in FRB~121102 \citep{scholz2016,gajjar2018}. Alternatively, the hypothesized central compact object could be producing bursts that do not cross our line-of-sight. To ensure our search was not affected by self-absorption, future observations of \first at higher radio frequencies are required.

The potential connection of FRB~121102 with long GRBs or SLSNe has sparked targeted searches for millisecond-duration bursts and compact persistent radio sources at the positions of such events. In one such a search, \cite{eftekhari2019} discovered a persistent radio source coincident with the SLSN PTF10hgi. An orphan GRB afterglow is explored as the potential origin of the emission, but is considered unlikely due to the high inferred isotropic jet energy, exceeding that of most observed long GRBs. Since the discovery of both this radio source and \first were motivated by observations of FRB~121102 and its environment, we compare the inferred isotropic jet energy for both sources. The inferred properties of the radio source associated with PTF10hgi, assuming GRB jet expansion, is estimated as $E_{\rm iso} \sim (3$--$5) \times 10^{53}\,\mathrm{erg}$, $n = 10^{-3}$--$10^{2}\,\mathrm{cm^{-3}}$ \citep{eftekhari2019}. Although this energy range is larger than the majority of observed long GRBs, it is comparable to that of \first ($E_{\mathrm{iso}} = 2 \times 10^{53}\,\mathrm{erg}$, $n = 10\,\mathrm{cm^{-3}}$; \citealt{law2018}). The results shown in the work presented here support the scenario in which \first is an orphan GRB afterglow. Whether this is the case for PTF10hgi as well is not clear at this point, but we argue that the inferred high isotropic jet energy in itself does not exclude an off-axis jet origin. The ultimate probe of this scenario is very high angular resolution VLBI observations. Accurately measuring the source size at the redshift of PTF10hgi (about five times more distant than \first) -- and especially considering its low flux density of $\sim 50\,\mathrm{\upmu Jy}$ -- is very challenging, and may only be possible with a very sensitive future SKA-VLBI array \citep{paragi2015} observing at high frequencies ($\gtrsim 5\,\mathrm{GHz}$).

\section{Conclusions} \label{sec:conclusions}

\first was reported as a slowly fading radio transient source. We provide the first constraints on the source size, using EVN data. These measurements confirm the non-thermal emission of the source and are consistent with jet expansion from a putative orphan long GRB. The derived average expansion velocity is consistent with a mildly relativistic expansion, noting that a significant deceleration has likely happened during these $\sim 30\,\mathrm{yr}$ after the event. A flux density lower than expected is reported, suggesting a faster decline after 2015. This decay could be explained by a change in the post-shock microphysical parameters following the transition to the non-relativistic phase, or by a drop in the ISM density (e.g.\ due to the shock reaching the outer edge of the star-forming region where the GRB exploded). We exclude scintillation-induced fluctuations as the origin of the reported variability.

Finally, although \first was discovered in a search for persistent radio sources similar to that associated with FRB~121102, we note significant differences between these sources (e.g. \first shows a significantly larger extend, and stronger luminosity decay). Still, \first could be a site of potential FRB production, although the burst searches conducted during the EVN observation reported null results.
Future radio observations are required to provide better constraints on the possible presence of FRBs arising from this object, as well as to characterize the evolution of the light-curve and its accelerated decay.

\section*{Acknowledgements}

We thank the annonymous referee for the useful comments and suggestions which helped to improve the paper.
We thank H.~J.~van~Langevelde for an internal review of the manuscript. K.N. would like to thank R.~A.~M.~J.~Wijers for helpful comments and discussions.
O.S. wishes to thank G.~Ghirlanda for valuable discussions and insights. The European VLBI Network is a joint facility of independent European, African, Asian, and North American radio astronomy institutes. Scientific results from data presented in this publication are derived from the following EVN project code: RM015.
e-MERLIN is a National Facility operated by the University of Manchester at Jodrell Bank Observatory on behalf of STFC.
We thank the directors and staff of all the EVN telescopes for making this target of opportunity observation possible.
We thank the staff of the Effelsberg Radio Telescope, and in particular U.~Bach, for his support with simultaneous pulsar recording.
B.M. acknowledges support from the Spanish Ministerio de Econom\'ia y Competitividad (MINECO) under grants AYA2016-76012-C3-1-P and MDM-2014-0369 of ICCUB (Unidad de Excelencia ``Mar\'ia de Maeztu'').
J.W.T.H. acknowledges funding from an NWO Vidi fellowship and from the European Research Council under the European Union's Seventh Framework Programme (FP/2007-2013) / ERC Starting Grant agreement nr. 337062 (``DRAGNET'').
E.P. acknowledges funding from an NWO Veni Fellowship.
R.K. is supported by the ERC synergy grant ``BlackHoleCam: Imaging the Event Horizon of Black Holes'' (Grant No. 610058).
This research made use of APLpy, an open-source plotting package for Python hosted at \url{http://aplpy.github.com}, Astropy, a community-developed core Python package for Astronomy \citep{astropy2013}, and Matplotlib \citep{hunter2007}.


\end{document}